\documentclass[11pt]{IEEEtran}
\usepackage{amssymb,amsmath}
\usepackage{subcaption}
\usepackage{graphicx}
\usepackage{eso-pic}
\usepackage[top=0.7in, left=0.65in, textheight=9.5in, textwidth=7.2in]{geometry}

\AddToShipoutPicture*{\footnotesize\sffamily\raisebox{0.3in}{\hspace{1.7cm}\fbox{\parbox{\textwidth}{\copyright~2015
IEEE. Personal use of this material is permitted. Permission from IEEE
must be obtained for all other uses, in any current or future media,
including reprinting/republishing this material for advertising or
promotional purposes, creating new collective works, for resale or
redistribution to servers or lists, or reuse of any copyrighted
component of this work in other works.}}}}

\title{Isogeometric Analysis Simulation of TESLA Cavities Under Uncertainty}

\author{%
	\IEEEauthorblockN{Jacopo Corno\IEEEauthorrefmark{1}\IEEEauthorrefmark{2}, Carlo de Falco\IEEEauthorrefmark{2}\IEEEauthorrefmark{3}, Herbert De Gersem\IEEEauthorrefmark{1}, Sebastian Sch\"ops\IEEEauthorrefmark{1}}\\%
	\IEEEauthorblockA{\IEEEauthorrefmark{1}Institut f\"ur Theorie Elektromagnetischer Felder, TU Darmstadt, Germany}\\%
	\IEEEauthorblockA{\IEEEauthorrefmark{2}MOX Modeling and Scientific Computing, Politecnico di Milano, Italy}\\%
	\IEEEauthorblockA{\IEEEauthorrefmark{3}CEN - Centro Europeo di Nanomedicina, Milano, Italy}%
}

\begin{document}
\maketitle

\begin{abstract}
In the design of electromagnetic devices the accurate representation of the geometry plays a crucial role in determining the device performance. For accelerator cavities, in particular, controlling the frequencies of the eigenmodes is important in order to guarantee the synchronization between the electromagnetic field and the accelerated particles.

The main interest of this work is in the evaluation of eigenmode sensitivities with respect to geometrical changes using Monte Carlo simulations and stochastic collocation. The choice of an IGA approach for the spatial discretization allows for an exact handling of the domains and their deformations, guaranteeing, at the same time, accurate and highly regular solutions.
\end{abstract}

\section{Introduction}

The performance of electromagnetic devices such as, for example, energy transducers, magnetrons, waveguides, antennas and linear accelerators is strongly related to the shape of the devices themselves. In particle accelerator cavities, in particular, the acceleration of the particle beam is achieved by exciting specific eigenmodes in the cavity and the resonant frequencies need to be synchronized with the flying particles to guarantee the acceleration. Even small mechanical deformations, either due to manufacturing imperfections or to the electromagnetic pressure (Lorentz detuning) may cause a non-negligible frequency shift \cite{Deryckere_2012aa,Corno_2014aa}.

A correct representation and handling of the domain is then of great importance when implementing a simulation scheme. It has been shown in \cite{Corno_2014aa} that the Isogeometric Analysis (IGA) discretization methods introduced in \cite{Hughes_2005aa, Buffa_2010aa} can produce a highly accurate solution, for example for the coupled electromagnetic-mechanic problem modeling Lorenz detuning.

The focus of this paper is on quantifying the uncertainty of linear accelerator superconducting cavities in frequency domain, or more precisely the sensitivity of the solution of Maxwell's eigenvalue problem 
\begin{equation}\label{eq:maxwell}
	\nabla\times\left(\mu^{-1}\nabla\times\mathbf{E}\right)=\omega^2\varepsilon\mathbf{E}
\end{equation}
with respect to (randomly) perturbed domains $\Omega(\mathbf{y})$, as in e.g. \cite{Xiao_2007aa}, where $\mathbf{E}$ is the electric field, $\mathbf{y}$ are geometry parameters, $\mu$, $\varepsilon$ and $\omega$ are the permeability, permittivity and angular frequency, respectively.
 
The paper is structured as follows, Section 2 and 3 discuss the main ideas behind IGA and Uncertainty Quantification (UQ). In Section 4 and 5 the methods are applied to a benchmark example (pillbox) and the TESLA cavity \cite{Aune_2000aa}. The sensitivity of the eigenfrequencies are investigated with respect to uncertain design parameters. Furthermore, non axis-symmetric deformations are taken into consideration.


\begin{figure}[b]
\vspace{-1em}
\centerline{\includegraphics[width=.6\columnwidth]{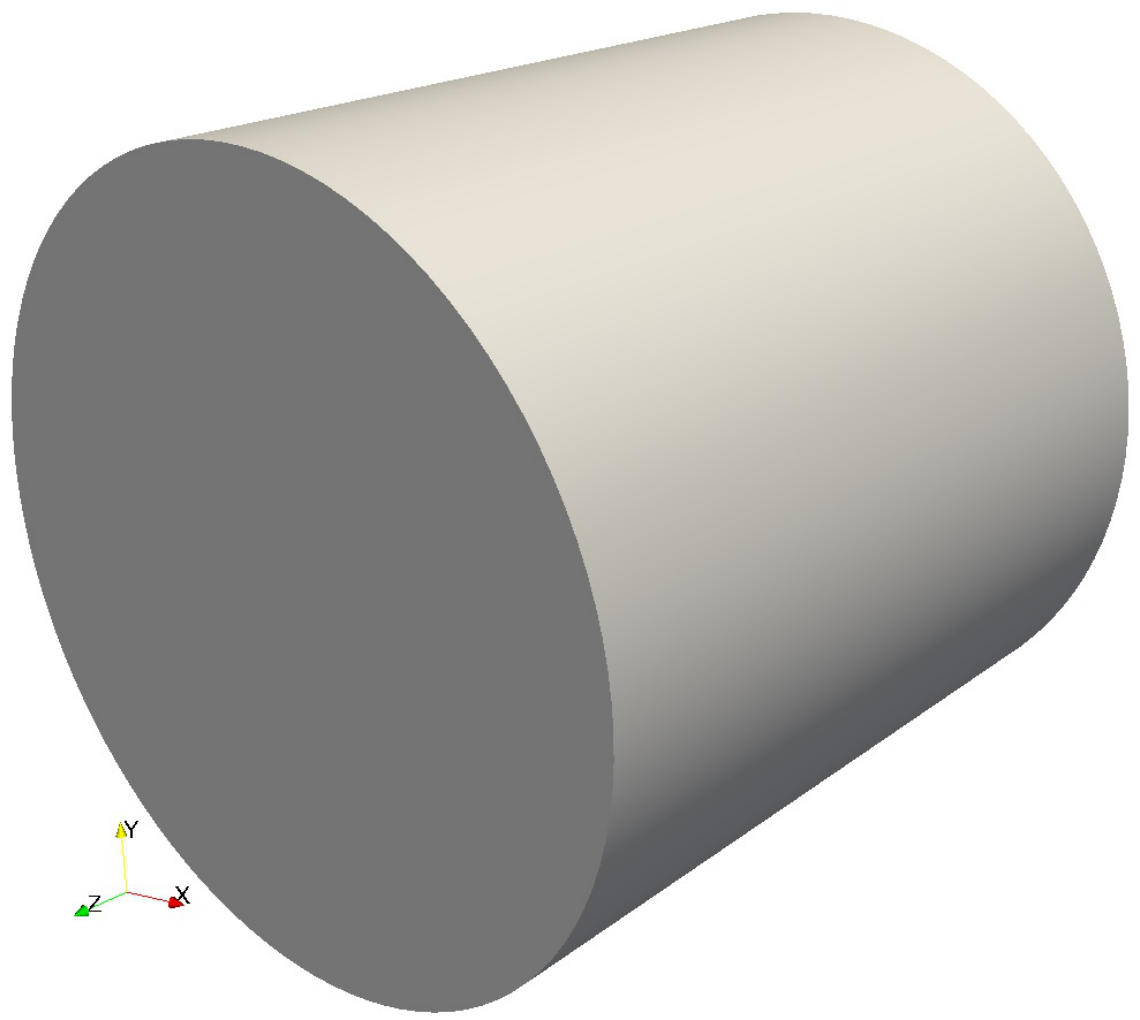}}
\caption{\small Cylindrical pillbox cavity.\label{fig:geo_pillbox} }\end{figure}

\section{Isogeometric Analysis}

In classical discretization methods such as the Finite Element Method (FEM), two steps are usually required: first, the meshing step, i.e., the discretization of the geometry from a Computer Aided Design (CAD) representation, in, typically,  hexahedral or tetrahedral elements. Secondly the discretization of the set of equations describing the problem to solve. In \cite{Hughes_2005aa}, Hughes et. al proposed a new approach, called Isogeometric Analysis, were the same basis functions that CAD uses for the description of geometries (e.g. B-Splines and NURBS) are used also as a basis for the solution of the partial differential equations. The IGA paradigm guarantees the exact description of the computational domain throughout all the analysis, even at the coarsest level of refinement.

$d$-dimensional B-Splines basis functions are defined in the reference space $[0,1]^\mathrm{d}$ following a tensor product approach. Along each dimension let $p$ be the degree and $\Xi=[\xi_{0},...,\xi_{n+p+1}]$ be a knot vector subdividing the unit segment  $[0,1]$. The Cox-de Boor formula 
then defines the $n+1$ basis functions $\{B_i^p\}_0^n$. A B-spline in the physical space is obtained by a mapping
\begin{equation}\label{eq:B-Spline_map}
\mathbf{x} = \sum_{i=0}^nB_i^p\mathbf{P}_i
\end{equation}
where $\mathbf{P}_i$ are called \emph{control points} and acts as degrees of freedom for the curve. Surfaces and volumes are the result of tensorization.

\begin{figure}[t]
\centerline{\includegraphics[width=.6\columnwidth]{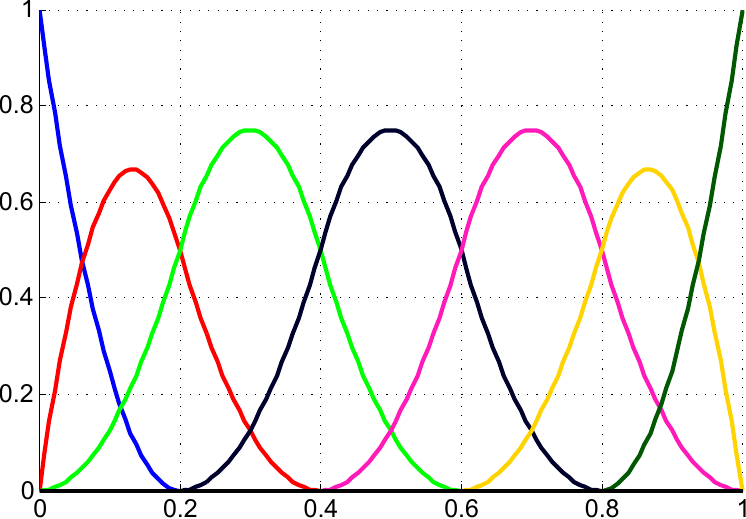}}
\caption{\small B-Spline basis functions of 2nd degree in 1D obtained with knots  $\Xi=[0,0,0,0.2,0.4,0.6,0.8,1,1,1].$\label{fig:B-Spline_basis} }
\vspace{-0.5em}
\end{figure}

As shown in Fig.~\ref{fig:B-Spline_basis}, B-Splines basis functions have higher regularity properties with respect to classical FEM (a B-Spline of degree $p$ can have in general up to $p-1$ global derivatives whereas FEM basis are only $C^0$ across the element boundaries). As a consequence of this, IGA has proven to have a higher accuracy w.r.t. the number of degrees of freedom than classical FEM.

The features of IGA are particulary beneficial for quantification of uncertainty  related to shape. The CAD representation is not lost in the meshing process, it is possible to easily modify the geometry and for each modified domain there is no need to perform remeshing when deforming the geometry.

\section{Uncertainty Quantification}

When dealing with mathematical models, e.g. partial differential equations, one has to consider that the input parameters might be affected by uncertainty. Let $\theta$ be a random event and $\mathbf{y}(\theta)$ a vector containing $N$ random input parameters $y_i(\theta)$ with density $\rho(\mathbf{y})$. Supposing the problem under consideration, i.e., \eqref{eq:maxwell}, is well defined for any $\mathbf{y}$, then the solution is itself a random variable
\begin{equation}\label{eq:u(y)}
\mathbf{E} = \mathbf{E}(\mathbf{y}(\theta)).
\end{equation}
The question is then how the input uncertainty $\mathbf{E}$ affects the quantity of interest $f(\mathbf{E})$, e.g., a cavity's eigenfrequency. In many applications classical (local) sensitivities, i.e., 
\begin{equation}
	\mathbf{D}_f(\mathbf{E})=\frac{\partial}{\partial \mathbf{E}}f(\mathbf{E})
\end{equation}
are a sufficient measure but often one is interested in statistics of outputs. They can be quantified by stochastic moments as expected value and variance
\begin{align}
	\mathbb{E}(f)&=\int_{-\infty}^\infty f(\mathbf{E}(\mathbf{y})) \rho(\mathbf{y})\,\mathrm{d}\mathbf{y}\\
	\mathrm{var}(f)&=\int_{-\infty}^\infty \big(f(\mathbf{E}(\mathbf{y}))-\mathbb{E}(f)\big)^2 \rho(\mathbf{y})\,\mathrm{d}\mathbf{y}
\end{align}
or the standard deviation $\mathrm{std}(f_0):=\sqrt{\mathrm{var}(f_0)}$. However, those integrals can rarely be solved exactly and thus one relies on numerical methods. 

One way out is the well-known Monte Carlo (MC) sampling method \cite{Liu_2002aa}. In this case the set of equations describing the problem is solved $M$ times, for $M$ realizations $\mathbf{y}_i$. The results obtained for $\mathbf{E}$ are then used to estimate expectations values of the desired quantity of interest by sample averages:
\begin{equation}
\mathbb{E}(f) \approx \frac{1}{M}\sum_{i=1}^{M}f(\mathbf{E}(\mathbf{y}_i)).
\end{equation}

More sophisticated approaches exploit the regularity of the solution to increase the speed of convergence, e.g., in the generalized Polynomial Chaos (gPC) the mapping (\ref{eq:u(y)}) from parameter to solution space is approximated by polynomials of degree $w$ \cite{Xiu_2010aa}. 
This approximation can be constructed by a basis of distribution dependent orthonormal polynomials $\{\psi_p(\mathbf{y})\}_0^w$ and a grid of points $\mathbf{y}_p$ in the parameter space (called collocation points), i.e.,
\begin{equation}
  \mathbf{E}(\mathbf{y}) \approx \sum_{p=0}^w\mathbf{E}(\mathbf{y}_p)\psi_p(\mathbf{y}).
\end{equation}
If the exact solution has a sufficiently smooth dependency w.r.t. its parameters this method converges exponentially. Unfortunately the rate depends heavily on the number of random parameters such that the advantage is lost for large $N$.

\begin{figure*}
  \begin{subfigure}[b]{.36\linewidth}
    \centerline{\includegraphics[width=0.98\columnwidth]{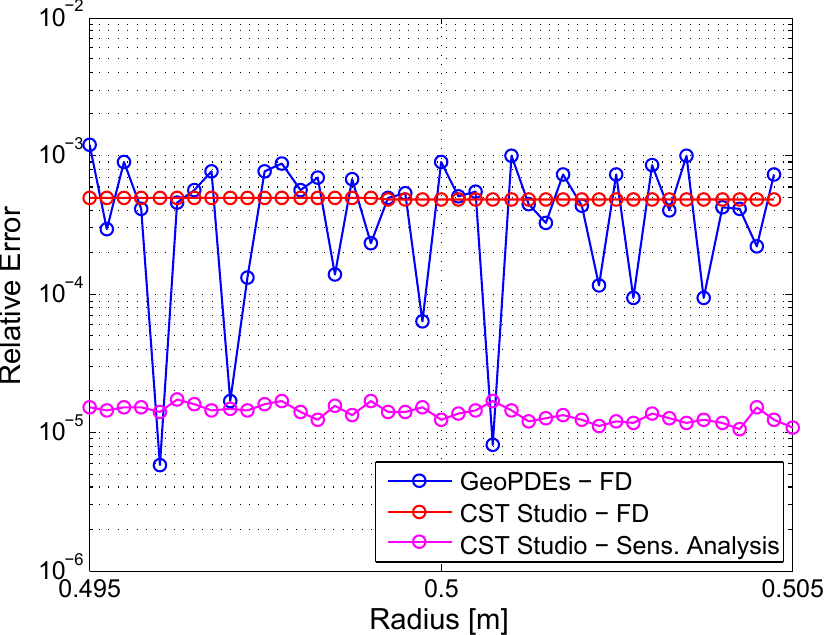}}
    \caption{\small Sensitivity of the fundamental frequency $f_0$ in the cylindrical cavity w.r.t. a change in the radius.\label{fig:mesh_sensitivity}}
  \end{subfigure}
  ~
  \begin{subfigure}[b]{.36\linewidth}
    \centerline{\includegraphics[width=.96\columnwidth]{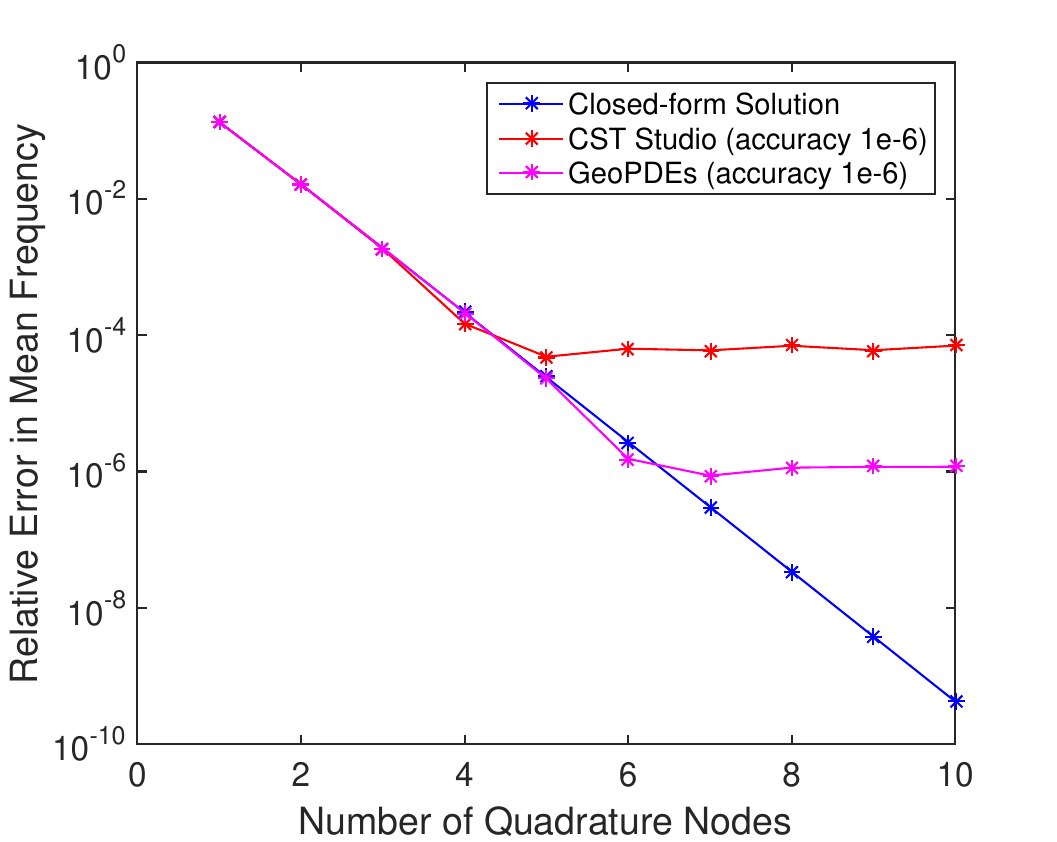}}
    \caption{\small Convergence of $\mathbb{E}(f_0)$ using gPC, i.e., stochastic Gauss quadrature, for various models.\label{fig:pillbox_convergence} }
  \end{subfigure}
  ~
  \begin{subfigure}[b]{.22\linewidth}
    \centerline{\includegraphics[width=\columnwidth]{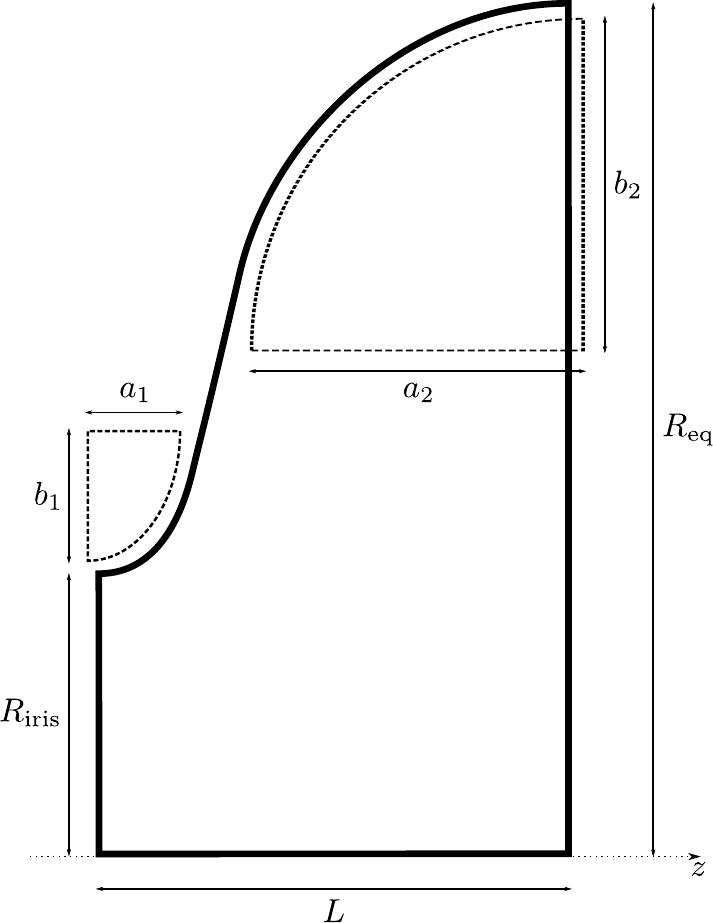}}
    \caption{\small Design parameters of the TESLA cavity half-cell \cite{Aune_2000aa}.
		\label{fig:TESLA_design} }
	\end{subfigure}
	\caption{Pillbox sensitivities, convergence of stochastic quadrature and TESLA cavity design}
\end{figure*}

\section{Pillbox Cavity}
As a benchmark example we consider the case of a cylindrical pillbox cavity with uncertain radius $r$ (see Fig.~\ref{fig:geo_pillbox}). We are interested in the resonant frequency of the fundamental (accelerating) mode $f_0$ and its sensitivity w.r.t. the change in radius close to the design value $r_{\mathrm{d}}=0.5$ m. 
In this settings one may exactly characterize the frequency and its derivative as
\[
f_0(r) = \frac{Gc}{2\pi r},\qquad \frac{\mathrm{d}f_0(r)}{\mathrm{d}r} = - \frac{Gc}{2\pi r^2}
\]
where $G\approx2.405$ is the first zero of the Bessel function of order $0$ and $c$ is the speed of light. In Figure~\ref{fig:mesh_sensitivity} we compare implementations of IGA (GeoPDEs~\cite{deFalco_2011aa}) to FEM (CST EM Studio~\cite{CSTEMStudio}). First, we compute the fundamental frequency for small perturbation of the radius across the design value, using second order IGA and second order FEM with the same level of accuracy. Secondly we use Finite Differences (FD) to estimate the sensitivity. It is clear from the figure that the FEM approximation suffer heavily from the need of remeshing the geometry each time the radius changes, while, even for this na\"ive approach, IGA obtains a smooth solution, since the parametrization of the cylinder doesn't change. Finally, the magenta curve in Fig.~\ref{fig:mesh_sensitivity} shows the results obtained by the automated sensitivity analysis tool available in CST: by using more sophisticated methods CST can achieve higher precision but oscillations are still present.

To study the global sensitivity of $f_0(r)$ an (artificial) random radius $r\sim\mathcal{U}(0.2\,\mathrm{m},0.8\,\mathrm{m})$ is considered. We compare the numerical approach (gPC with FEM and IGA) to the analytical solution
\begin{align*}
	\mathbb{E}(f_0)&=0.2651115...~\textrm{Hz},\\ 
	\mathrm{std}(f_0)&=0.1095555...~\textrm{Hz}.
\end{align*}
Fig.~\ref{fig:pillbox_convergence} depicts the rapid convergence until the accuracy of the spatial discretization is reached.

\begin{figure}[t]
\vspace{-.3cm}
\centerline{\includegraphics[width=.8\columnwidth]{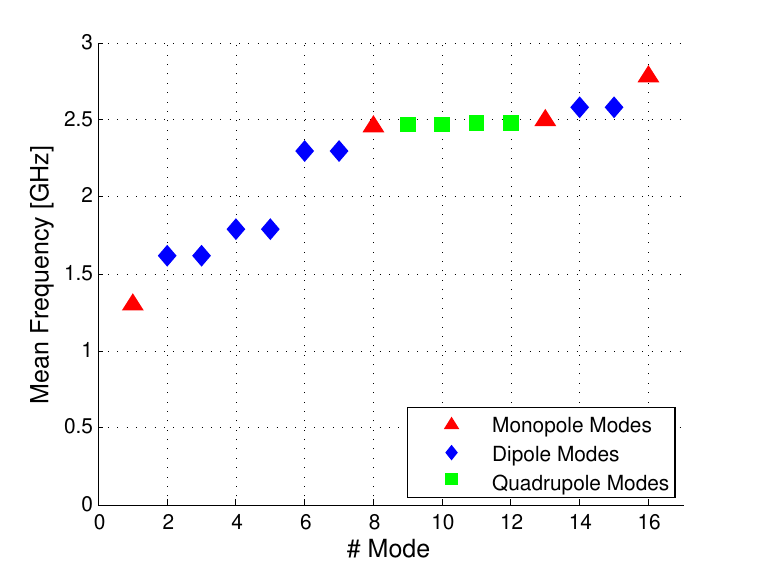}}
\caption{First 16 modes of one-cell TESLA cavity.\label{fig:UQ_design_f}}
\end{figure}

\begin{figure*}[t]
  \begin{subfigure}{.49\linewidth}
  	\centering
    \includegraphics[width=0.75\columnwidth]{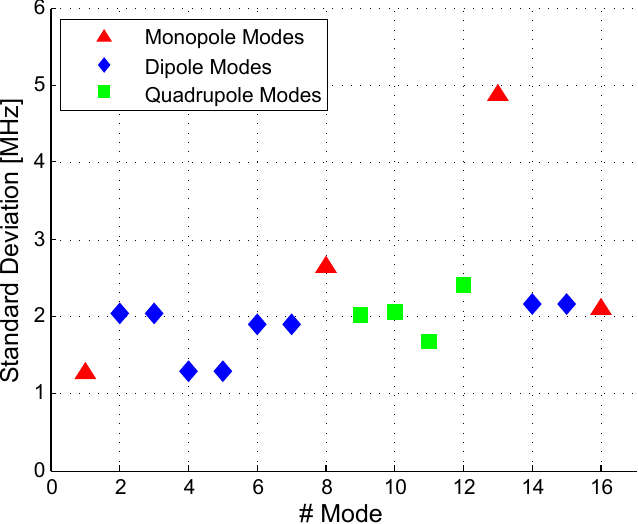}
    \caption{Eigenmodes due to uncertain design parameters\label{fig:UQ_design_f_sigma}}
  \end{subfigure}
  ~
  \begin{subfigure}{.49\linewidth}
    \centering
    \includegraphics[width=0.75\columnwidth]{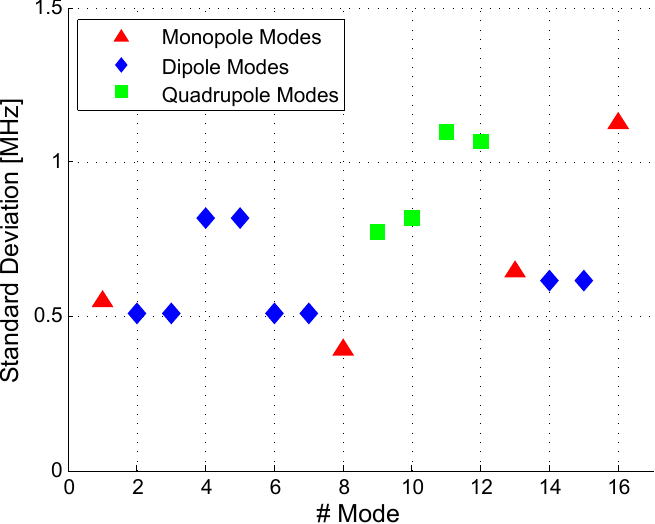}
    \vspace{1mm}
    \caption{Eigenmodes due to elliptic deformation\label{fig:UQ_elliptic_f_sigma}}
  \end{subfigure}
\caption{\small Standard deviations for the first 16 modes of the one-cell TESLA Cavity.}
\end{figure*}

\section{TESLA Cavity}

Let us now consider the single cell TESLA cavity geometry whose seven design parameters are described in Figure~\ref{fig:TESLA_design}.  Following~\cite{Schmidt_2014aa}, we apply UQ for design parameters albeit this restricts possible deviations significantly and more realistic cases, such as non-axissymmetric deformations, bumps/kinks due to welding, mechanical deformation due to Lorentz forces or misalignment of the irisis cannot be represented. 

We consider the parameters to be uncertain with mean values equal to the TESLA mid-cell design~\cite{Aune_2000aa} and deviations $y_i\sim\mathcal(-0.125\,\mathrm{mm},0.125\,\mathrm{mm})$. MC sampling is  applied to estimate mean values and standard deviation of the resonant frequencies for the first 16 modes. 

With respect to~\cite{Schmidt_2014aa}, where a 2D solver was used to compute the accelerating frequency by exploiting the cylindrical symmetry of the geometry, GeoPDEs is able to solve the full 3D cavity, thus allowing for the computation of the full spectrum of frequencies, Fig. \ref{fig:UQ_design_f}. In Fig.~\ref{fig:UQ_design_f_sigma} the standard deviation of these modes are depicted. The values are in the MHz range with a maximum deviation in correspondence of the third monopole mode.

A second case investigates elliptic deformations of the cavity. For this the domain is deformed in such a way that the cross section is no more a circle but an ellipse, i.e., breaking symmetry. The deviation with respect to the design geometry is chosen to be a uniformly distributed random variable with zero mean and $\sigma = 6.6667\cdot10^{-5}$, i.e.,  $R_{\mathrm{eq}}\sim\mathcal{U}(103.1\mathrm{mm},103.5\mathrm{mm})$. The solution of the eigenproblem is computed on a $10\times10$ grid of collocation points for the first 16 modes as in the previous case. Results (in Fig.~\ref{fig:UQ_elliptic_f_sigma}) suggest that possible elliptical deformations of the cavity play a slightly weaker role than the design parameters. Nevertheless the higher order modes are, once again, more effected than the fundamental one. Future studies on the full 9-cell cavity should take into account the necessity of including these modes in the analysis.

\section{Conclusions}
In this paper the usage of IGA and gPC for the uncertainty quantification of cavities was proposed since those methods exploit the smoothness of the solution in spatial and parameter domain. Academic and realistic geometries underline that this methodology allows to accurately quantify the impact of uncertainties.

\section*{Acknowledgments}
{\footnotesize
This work is supported by the ``Excellence Initiative'' of the German Federal and State Governments. the Graduate School CE at TU Darmstadt and the DFG network ``UQ for Cavities'' (SCHM3127/2-1). C. de Falco's work is partially funded by the ``Start-up Packages and PhD Program project'', co-funded by Regione Lombardia through ``Fondo per lo
sviluppo e la coesione 2007-2013'' (FAS program).

}


\begin{thebibliography}{9}
\itemsep=0ex
  \bibitem{Xiao_2007aa} L. Xiao et al., \emph{Modeling imperfection effects on dipole modes in TESLA cavity}, IEEE PAC 2007, 2454--2456, 2007.
  \bibitem{Deryckere_2012aa} J. Deryckere, T. Roggen, B. Masschaele, and H. De Gersem, \emph{Stochastic response surface method for studying microphoning and lorentz detuning of accelerators cavities}, ICAP2012, 158-160, 2012.
	\bibitem{Hughes_2005aa} T. J. R. Hughes, J. A. Cottrell, and Y. Bazilevs. \emph{Isogeometric analysis: CAD, finite elements, NURBS, exact geometry and mesh refinement}, Comp. Meth. in App. Mech. and Eng., 194(39), 4135-4195, 2005.
	\bibitem{Buffa_2010aa} A. Buffa, G. Sangalli, and R. Vázquez, \emph{Isogeometric analysis in electromagnetics: B-splines approximation}, Comp. Meth. in App. Mech. and Eng., 199, 1143-1152, 2010.
	\bibitem{Liu_2002aa}Liu, J.S., \emph{Monte Carlo Strategies In Scientific Computing}, Harvard University, 2002.
	\bibitem{Aune_2000aa} B. Aune et al., \emph{Superconducting tesla cavities. Physical Review Special Topics - Accelerators and Beams}, 3(9):092001, 2000. 
	\bibitem{Corno_2014aa} J. Corno, C. de Falco, H. De Gersem and S. Sch\"ops, \emph{Isogeometric Simulation of Lorentz Detuning in Superconducting Accelerator Cavities}, MOX Report 31/2014.
	\bibitem{deFalco_2011aa} C. de Falco, A. Reali, R. V\'{a}zquez, \emph{GeoPDEs: A research tool for Isogeometric Analysis of PDEs}, Advances in Eng. Soft., 12(42), 1020-1034, 2011.
	\bibitem{Xiu_2010aa} D. Xiu, \emph{Numerical Methods for Stochastic Computations: A Spectral Method Approach}, Princeton University Press, 2010.
	\bibitem{CSTEMStudio} {CST EM STUDIO\textregistered}, Computer Simulation Technology AG, \texttt{www.cst.com}.
	\bibitem{Schmidt_2014aa} C. Schmidt, T. Flisgen, J. Heller, U. van Rienen, \emph{Comparison of techniques for uncertainty quantification of superconducting radio frequency cavities}, ICEAA 2014, 117--120, 2014.
	\bibitem{Gravesen_2011aa} J. Gravesen, A. Evgrafov und D.-M. Nguyen, \emph{On the sensitivities of multiple eigenvalues}, Structural and Multidisciplinary Optimization 44.4, 583–-587, 2011.
\end{thebibliography}
\end{document}